\documentclass[12pt]{article}
\usepackage{amsmath}
\usepackage{amsfonts}
\usepackage{amssymb}
\usepackage{graphicx}%
\providecommand{\U}[1]{\protect\rule{.1in}{.1in}}
\begin{document}

\begin{center}
{\Large {\bf Axion, Neutrinos Masses and $\mu$-Problem in Minimal Supersymmetric Standard Model (MSSM)}}\\
M. C. Rodriguez   \\
{\it Grupo de F\'\i sica Te\'orica e Matem\'atica F\'\i sica \\
Departamento de F\'\i sica \\
Universidade Federal Rural do Rio de Janeiro - UFRRJ \\
BR 465 Km 7, 23890-000 \\
Serop\'edica, RJ, Brazil, \\
email: marcoscrodriguez@ufrrj.br \\} 
\end{center}

\date{26/06/2020}

\begin{abstract}
We review very nice mechanism to generate masses to all the 
neutrinos in the Minimal Supersymmetric Standard Model (MSSM) 
without break $R$-parity. In consequence 
we get viable axion as Dark Matter candidate and at the same time 
solve the $\mu$-problem in a similar way as done in the Next to 
Minimal Supersymmetric Model (NMSSM). 
\end{abstract}

PACS number(s): 12.60. Jv


\section{Introduction}

Two  important problems  of today's particle physics are: 
\begin{itemize}
\item Explain the  neutrino masses and their oscillations parameters;
\item To solve the strong CP-problem and get an axion as viable candidate 
to be Dark Matter. 
\end{itemize}
Certainly the most popular extension of the Standard Model (SM) 
is its supersymmetric counterpart called Minimal Supersymmetric 
Standard Model (MSSM) 
\cite{Rodriguez:2009cd,Rodriguez:2016esw,Rodriguez:2019,
Rodriguez:2020,dress,Baer:2006rs,ait1,wb}. 
The main motivation to study this model are:
\begin{itemize}
\item We get unification of the three  gauge couplings of Standard Model;
\item Solve the hierarchy Problem (why $M_{H}\ll M_{P}$);
\item It is possible to understand the electroweak symmetry breaking.
\end{itemize}

Supersymmetric theories (SUSY) has also made several correct predictions \cite{Chung:2003fi}:
\begin{itemize}
\item SUSY predicted in the early 1980s that the top quark would be heavy;
\item SUSY GUT theories with a high fundamental scale accurately predicted the present experimental value of $\sin^{2} \theta_{W}$ 
before it was mesured;
\item SUSY requires a light Higgs boson to exist.
\end{itemize}
Together these success provide powerful indirect evidence that low energy SUSY is indeed part of correct description of nature. 

The MSSM, however, suffers from the $\mu$-problem.  It means that the predictions of the MSSM set a phenomenological constraint on 
the magnitude  $\mu \sim {\cal O}(M_{W})$, we use this fact in order to generate masses to two neutrinos when we broke $R$-parity in 
the MSSM \cite{hall,banks,Romao:1991ex,rv1,Davidson:2000ne,Montero:2001ch,
Smirnov:2004hs}, while the natural cut-off scale is the Planck scale $M_{P} \sim 10^{19}\,\ GeV$. The MSSM 
does not provide any mechanism to explain the difference between the two scales. This is know as the \emph{$\mu$ problem} 
\cite{Rodriguez:2016esw,Rodriguez:2019,dress,Baer:2006rs,ait1}.

The simplest way to solve the $\mu$-problem is to introduce the Next-to-the-Minimal Supersymmetric Standard-Model (NMSSM) \cite{Rodriguez:2016esw,Rodriguez:2019,dress,Baer:2006rs,nmssm}. We add to the usual 
supermultiplets a new   scalar singlet field (added in a chiral supermultiplet) \cite{Rodriguez:2016esw,Rodriguez:2019}
\begin{equation} 
\hat{S}\sim({\bf 1},0),
\label{singletnmssm}
\end{equation}
where this chiral superfield is expanded as \cite{wb}
\begin{eqnarray}
\hat{S}(y, \theta )&=&S(y)+ \sqrt{2}\theta \tilde{S}(y)+ \theta \theta F_{S}(y), 
\label{singletoquebrapq}
\end{eqnarray}
where $S$ is the scalar in the singlet responsable to generate the 
$\mu$ term when this new field develops its VEV 
\cite{Rodriguez:2019,dress}. Another essential feature of the 
NMSSM is the fact that the mass bounds for the Higgs bosons and 
neutralinos are weakened constrained when compared to those in the 
MSSM. For more details about the scalar sector of this model see Ref.~\cite{Rodriguez:2019,dress,Baer:2006rs,Drees:1988fc,
Ellis:1988er}. This model has also good results in cosmology as presented 
in the followings articles \cite{Rodriguez:2019,dress,Baer:2006rs}. 

The strong CP problem arise as in the following way, we will 
consider Quantum Chromo-Dinamycs (QCD) with 
one flavor of massless quark, given by $\Psi$. The path integral that describes this quark, denoted as $\Psi$, propagation and its 
interaction with the gluons, $\left(G^{a} \right)_{m}$, 
is \cite{Srednicki:2004hf}
\begin{equation}
Z= \int {\cal D}A {\cal D}\Psi {\cal D}\bar{\Psi} \exp \left\{
\imath \int d^{4}x \left[ \imath \bar{\Psi} \left( D^{m}\gamma_{m}\right) \Psi - 
\frac{1}{4}\left(G^{a}\right)^{mn}
\left(G^{a}\right)_{mn}- \frac{g^{2}_{s}\theta}{32 \pi^{2}}
\epsilon_{mnop} \left(G^{a}\right)_{mn} \left(G^{a}\right)_{op} \right] \right\},
\end{equation}
where the covariant derivate is defined as
\begin{equation}
{\cal D}_{m}= \partial_{m}+ \imath g_{s}\left(G^{a}\right)_{m}T^{a},
\label{covder}
\end{equation}
where $g_{s}$ is the strong coupling constant, $T^{a}$ are the generator of the group $SU(3)$, while $\left(G^{a}\right)_{m}$ is the gluonic field and 
the field-strength tensor to gluons are
\begin{equation}
\left( G_{a}\right)^{mn} = \partial^{m} \left(G_{a}\right)^{n} - \partial^{n} \left(G_{a}\right)^{m} + 
g_{s} f_{abc} \left(G_{b}\right)^{m} \left(G_{c}\right)^{n},
\label{strengthfieldgluons}
\end{equation}
the structure constant $f_{abc}$, of the gauge group $SU(3)_{C}$, is 
defined as
\begin{equation}
[T_{a},T_{b}]= \imath f_{abc}T_{c},
\label{fabc}
\end{equation}
are completely antisymmetric. 

The strong CP problem simply questions why the effective 
$\overline{\theta}$-term is so small, 
or in other words, what is the origin of such a strong 
fine tuning between the initial value $\theta$ and 
the phases of the quark mass matrices? 

Axions are fascinating hypothetical particles whose existence was proposed by S.
Weinberg and F. Wilczek to give a resolution of the strong CP problem \cite{ww}. The problem is resolved by invoking 
a dynamical mechanism \cite{pq} to relax the above $\theta_{QCD}$ parameter 
(including all contributions from colored fermions) to zero.  However, this 
necessarily results \cite{ww} in a very light pseudoscalar particle called 
the axion, which has not yet been observed \cite{exp4}.

The famous Peccei-Quinn  solution \cite{pq} is based on the concept 
of spontaneously broken global axial symmetry $U(1)_{\rm PQ}$.  
In the context of this mechanism $\overline{\theta}$  
essentially becomes a dynamical field defined as 
\begin{eqnarray}
\overline{\theta}= \frac{a}{f_{a}},
\end{eqnarray}
with an effective 
potential induced by the non-perturbative QCD effects  
which fixes vacuum expectation value (VEV) at zero.   
Here $a$ stands for the pseudo-Goldstone mode of the 
spontaneously broken PQ symmetry, an axion,  
and $f_{a}$ is 
a VEV of scalar (or a VEV combination of several scalars) 
responsible for the $U(1)_{\rm PQ}$ symmetry breaking. 
This scale is model dependent. 
In particular, 
in the original Weinberg-Wilczek (WW) model \cite{ww}  
$f_{a}$ is order electroweak scale. 

The main solutions to avoid the constraints in the axion are:
\begin{itemize}
\item The Dine-Fischler-Srednicki-Zhitnitskii (DFSZ) solution \cite{dfsz} introduces a heavy singlet scalar field as the 
source of the axion but its mixing with the doublet scalar 
fields (which couple to the usual quarks) is very much suppressed;  
\item The Kim-Shifman-Vainshtein-Zakharov (KSVZ) 
solution \cite{ksvz} also has a heavy singlet scalar field but 
it couples only to new heavy colored fermions.
\end{itemize}

In supersymmetric theories, we can introduce the anomalous 
interaction between axion with gluons\footnote{This term can be generated from string theories as shown at \cite{Conlon:2006tq}.} in the following way \cite{Higaki:2011bz,Bae:2014efa}:
\begin{equation}
\int d^{2}\theta \left( \hat{A} Tr[W^{\alpha}W_{\alpha}]\right) +hc \propto 
(s+ \imath a)Tr[G^{a}_{mn}G^{amn}],
\label{axiongluoncoupling}
\end{equation}
where the axion chiral superfield is defined as \cite{wb}
\begin{equation}
\hat{A}(y, \theta)= \frac{1}{\sqrt{2}}\left( s(y)+ \imath a(y) \right)+ \sqrt{2}\theta 
\tilde{a}(y)+ \theta \theta F_{a}(y),
\label{axionsuperfield}
\end{equation} 
where $s$ is the saxion while $\tilde{a}$ is the axino, $a$ is the axion field and 
$F_{a}$ is the auxiliary field of the axion supermultiplet. The axino and 
saxion remain also massless in the case where Supersymmetry is hold. Supersymmetry breaking 
induces their masses wchich are generically expected to be of order of Supersymmetry breaking 
scale. The field value of the saxion (corresponding to the axion decay constant)
needs to be stabilized, that requries a coupling to the SUSY breaking
sector.

In Eq.(\ref{axiongluoncoupling}), $a$ is colour indices and $G^{a}_{mn}$ is the gluonic field strength. This 
term will generate the following interaction between axion-gluon
\begin{equation}
g_{s}f^{abc}(s+ \imath a)G^{a}_{m}G^{b}_{n},
\end{equation}
where $g_{s}$ is the strong coupling constant, $f^{abc}$, defined at 
Eq.(\ref{fabc}), is the structure coefficients of 
$SU(3)_{C}$ and $G^{a}_{m}$ are the gluons.

There are some supersymmetrics models, where we introduce the axion, as the 
follows \cite{Bae:2014efa}:
\begin{itemize}
\item KSVZ: $\hat{\Phi}+ \hat{\Phi}^{c}$ is $3 \oplus \bar{3}$ heavy quarks
\begin{equation}
W= \lambda_{1}\hat{S}\hat{\Phi}\hat{\Phi}^{c}
\end{equation}
the axion solution is realized by the presence of extra  $\hat{\Phi}$ and $\hat{\Phi}^{c}$
\item DFSZ: $\hat{\Phi}+ \hat{\Phi}^{c}$ is $5 \oplus \bar{5}$ under $SU(5)$
\begin{equation}
W= \lambda_{1}\hat{S}\hat{\Phi}\hat{\Phi}^{c}+ \lambda_{2} 
\frac{\hat{S}\hat{S}}{M_{Pl}}\hat{H}_{1}\hat{H}_{2}.
\end{equation}
\end{itemize}
Therefore, we can easilly see, the $\mu$-problem of the MSSM is connected to the axion 
solution and it plays a major role in the axino/saxion cosmology. However, 
the neutrinos are still massless. 

Some years ago, it was shown that we can give masses to all neutrinos and at same time solve the $\mu$-problem and CP-problem, including more singlet fields in this
model~\cite{Ma:2001ac}. These new singlets, in this case, introduce a new energy 
scale, where the Peccei-Quinn symmetry is broken, which   allow  Majorana Mass  terms to 
the right handed neutrinos, but  avoids the term responsible  for the $\mu$-problem in the MSSM.

The outline of this review is the following: In Sec.(\ref{sec:massiveneutrinosaxion}) we discuss 
how to implement the axion and at same time generate masses to all neutrinos and also solve the $\mu$-problem and CP-Problem, 
we get the axion field in Sec.(\ref{sec:scalarpot}). In the end 
we present our conclusion.

\section{Neutrinos Massives with the Axion.}
\label{sec:massiveneutrinosaxion}

Here we will present briefly how we can use the axion and get 
mass to all neutrinos as presented at \cite{Ma:2001ac}, the quark sector 
is exact the same as defined at Standard Model. In 
order to generate masses to neutrinos beyond the usual leptons 
of the MSSM, we will introduce right-handed neutrinos, in similar 
way as it is done in the MSSM with three right-handed neutrinos 
(MSSM3RHN)~\cite{Rodriguez:2020,Baer:2006rs}. The new aspect now, we 
also associate to each superfield the following $PQ$-charges 
given in Tab.(\ref{allleptonscharges}). 

\begin{table}[h]
\begin{center}
\begin{tabular}{|c|c|c|c|}
\hline  
$\mbox{ Chiral \,\ Superfield}$ & $(B-L)$-charge & $R$-charge & $PQ$-charge \\
\hline 
$\hat{L}_{iL}\sim({\bf 1},{\bf 2},-1)$ & $-1$ & $+ \left( \frac{1}{2}\right)$ & $+ \left( \frac{1}{2}\right)$  \\ \hline
$\hat{E}_{iR}\sim({\bf 1},{\bf 1},2)$ & $+1$ & 
$- \left( \frac{1}{2}\right)$ & $+ \left( \frac{1}{2}\right)$  \\ \hline
$\hat{N}_{iR}\sim({\bf 1},{\bf 1},0)$ & $+1$ & $- \left( \frac{1}{2}\right)$ & $+ \left( \frac{1}{2}\right)$  \\ \hline
$\hat{H}_{1}\sim({\bf 1},{\bf 2},-1)$ & $0$ & $0$ & $-1$ \\ \hline 
$\hat{H}_{2}\sim({\bf 1},{\bf 2},1)$ & $0$ & $0$ & $-1$ \\ \hline
$\hat{S}_{0}\sim({\bf 1},{\bf 1},0)$ & $0$ & $0$ & $-2$ \\ \hline
$\hat{S}_{1}\sim({\bf 1},{\bf 1},0)$ & $0$ & $+1$ & $-1$ \\ \hline 
$\hat{S}_{2}\sim({\bf 1},{\bf 1},0)$ & $0$ & $0$ & $+2$ \\ 
\hline
\end{tabular}
\end{center}
\caption{\small Quantum number assignment to the leptons and new 
scalars of the model. The  families index for leptons are $i,j=1,2,3$. The parentheses are the transformation 
properties under the respective representation of $(SU(3)_{C},SU(2)_{L},U(1)_{Y})$.}
\label{allleptonscharges}
\end{table}

We will introduce three singlet superfields, to give Majorana 
masses to all neutrinos and solve also the $\mu$-problem
$\hat{S}_{0,1,2}$, whose quantum numbers are listed in Tab.~(\ref{allleptonscharges}). These  singlets will break 
the $PQ$-symmetry \cite{Ma:2001ac} and their vaccum expectation values are denoted as 
\begin{eqnarray}
\langle S_{0} \rangle &=& \frac{x_{0}}{\sqrt{2}}, \,\ 
\langle S_{1} \rangle = \frac{x_{1}}{\sqrt{2}}, \,\ 
\langle S_{2} \rangle = \frac{x_{2}}{\sqrt{2}}.
\end{eqnarray}
 
Using these new fields we can constructs the following supersymmetric lagrangians
\begin{eqnarray}
{\cal L}_{lepton}&=& \int d^{4}\theta\;\sum_{i=1}^{3}\left[\,
K\left( \hat{ \bar{L}}_{iL}e^{2[g\hat{W}+
g^{\prime} \left( -\frac{1}{2} \right) \hat{b}^{\prime}]}, \hat{L}_{iL} \right) +
K\left( \hat{ \bar{E}}_{iR} e^{2[g^{\prime} 
\left( \frac{2}{2} \right) \hat{b}^{\prime}]}, 
\hat{E}_{iR} \right)  
\right. \nonumber \\ &+& \left.
K(\hat{\bar{N}}_{iR},\hat{N}_{iR})\,\right] \,\ , \nonumber \\
{\cal L}_{Higgs}&=&  \int d^{4}\theta\;\left[\,
K\left( \hat{ \bar{H}}_{1}e^{2[g\hat{W}+
g{\prime} \left( \frac{-1}{2} \right) \hat{b}^{\prime}]}, \hat{H}_{1} \right) +
K\left( \hat{ \bar{H}}_{2}e^{2[g\hat{W}+
g^{\prime} \left( \frac{1}{2}
\right) \hat{b}^{\prime}]}, \hat{H}_{2}\right) \right.  \nonumber \\
&+& \left. 
K(\hat{\bar{S}}_{0},\hat{S}_{0})+ K(\hat{\bar{S}}_{1},\hat{S}_{1}) 
+ K(\hat{\bar{S}}_{2},\hat{S}_{2}) \right] 
+  \int d^{2}\theta\; W+ \int d^{2}\bar{\theta}\;\bar{W}\!, 
\label{allsusytermsm1}
\end{eqnarray}
where $W$ is the superpotential of this model, while $K$ is 
the K\"ahler potential and it is defined as \cite{wb}
\begin{eqnarray}
\int d^{4}\theta K( \hat{\bar{S}}, \hat{S})&\equiv& \hat{\bar{S}}\hat{S} =
- \left( \partial_{m}S^{\dagger} \right) \left( \partial^{m} S \right) + 
\imath \bar{\tilde{S}}\bar{\sigma}^{m} \partial_{m}\tilde{S}+ 
\bar{F}_{S}F_{S}, \nonumber \\
\int d^{4}\theta K( \hat{\bar{S}}e^{\imath gT^{a}\hat{V}^{a}}, \hat{S})&\equiv& 
\hat{\bar{S}}e^{gT^{a}\hat{V}^{a}}\hat{S} =- 
\left( {\cal D}_{m}S \right)^{\dagger}\left( {\cal D}^{m}S \right)- 
\imath \bar{\tilde{S}}\bar{\sigma}^{m} {\cal D}_{m}\tilde{S}+\bar{F}_{S}F_{S} 
\nonumber \\ &+&
\imath \sqrt{2}g \left[ S^{\dagger} \left(T^{a}\lambda^{a}\right) \tilde{S}- 
\overline{\tilde{S}} \left( \bar{\lambda}^{a}\bar{T}^{a}\right)S \right] 
+gD^{a}S^{\dagger}T^{a}S, \nonumber \\
\label{kahlerpotential}
\end{eqnarray}
where $\hat{S}$ is a chiral superfield, defined at 
Eq.(\ref{singletoquebrapq}), ${\cal D}_{m}$ is the covariant 
derivative defined at Eq.(\ref{covder}), while $g$ is the 
constant coupling of same gauge group and $\hat{V}$ is the vector 
superfield associated with this gauge group.

The last term in Eq.(\ref{allsusytermsm1}), $W$ is the superpotential of this model and it is given by
\begin{eqnarray}
W&=&m \hat{S}_{2}\hat{S}_{0}+f \hat{S}_{1} \hat{S}_{1}\hat{S}_{2}+
h^{H}\left( \hat{H}_{1}\hat{H}_{2}\right) \hat{S}_{2} +
f^l_{ij}\left( \hat{H}_{1}\hat{L}_{iL}\right) \hat{E}_{jR}
+f^{\nu}_{ij}\left( \hat{H}_{2}\hat{L}_{iL}\right) \hat{N}_{jR}+ 
f^{M}_{ij}\hat{N}_{iR}\hat{N}_{jR}\hat{S}_{1}. \nonumber \\ 
\label{totalsuperpotential}
\end{eqnarray}
These superpotential conserve $R$-parity and respect the $PQ$-symmetry. The $\mu$-term is forbidden and it is rewritten by 
$h^{H}$-coupling, in simillar way as done at 
NMSSM~\cite{Rodriguez:2016esw,Rodriguez:2019}.

The terms proportional to $f^{\nu}$ generate Dirac mass term $M^{D}=f^{\nu} \langle H_{2}\rangle$ and $f^{M}$
generate Majorana mass term as 
$M_{M}=f^{M} \langle S_{1}\rangle$, which is necessary to
generate the type I see-saw mechanism mass term for the active neutrinos, in similar way as happen at MSSM3RHN~\cite{Rodriguez:2020}.

\section{Scalar Potential}
\label{sec:scalarpot}

We can rewrite our potential in the 
following way
\begin{equation}
V=V_{MSSM}+V_{\phi}+V_{\phi-H}+V_{H-axion}+V_{axion},
\label{ep1}
\end{equation}
where
\begin{eqnarray}
V_{MSSM}&=&\left( m^{2}_{1}+ \frac{(h^{H})^{2}x^{2}_{2}}{2} \right) |H_{1}|^{2}+ 
\left( m^{2}_{2}+ \frac{(h^{H})^{2}x^{2}_{2}}{2} \right) |H_{2}|^{2}+
(h^{H})^{2}|H_{1}H_{2}|^{2} \nonumber \\ &+& 
\left( \frac{g^{2}+g^{\prime 2}}{8}\right) \left( |H_{1}|^{2}-|H_{2}|^{2}\right)^{2}+ 
\frac{g^{2}}{2}|H^{\dagger}_{1}H_{2}|, \nonumber \\
V_{H-axion}&=&m_{12}S_{2}H_{1}H_{2}+2h^{H}\left( mS^{*}_{0}+fS^{*2}_{1}\right) H_{1}H_{2}+hc
, \nonumber \\
V_{axion}&=&\mu^{2}_{0}|S_{0}|^{2}+ \mu^{2}_{1}|S_{1}|^{2}+ 
\left( \mu^{2}_{2}+ \frac{(h^{H})^{2}(v^{2}_{1}+v^{2}_{2})}{2} \right) |S_{2}|^{2}+ 
|mS_{0}+fS_{1}S_{1}|^{2} \nonumber \\
&+&m^{2}|S_{2}|^{2}+4f^{2}|S_{1}|^{2}|S_{2}|^{2}+[\mu_{20}^{2} S_{2} S_{0} + \mu_{112} S_{1}^{2} S_{2} +hc].
\end{eqnarray}
We have to analayse $V_{axion}$, given above, to get the axion. But we can make some supositions in order to simplify our analyses. 

We will first break the Peccei-Quinn symmetry but supersymmery is not broken \cite{Ma:2001ac}. 
In this case 
\begin{eqnarray}
V_{axion}&=&|mS_{0}+fS_{1}S_{1}|^{2}+m^{2}|S_{2}|^{2}+4f^{2}|S_{1}|^{2}|S_{2}|^{2}.
\end{eqnarray}

It is simple to show it has the trivial solution, to get the 
extremum in $V_{axion}$, is given by $x_{0}=x_{1}=x_{2}=0$. There 
are the second solution given by \cite{Ma:2001ac}:
\begin{equation}
x_{2}=0, \,\ x_{0}=- \frac{fx^{2}_{1}}{m}.
\label{getaxion}
\end{equation}
These values breaks $U(1)_{PQ}$ spontaneously. 

We will make the following shifts 
$\hat{S}_{2,1,0}\rightarrow \hat{S}_{2,1,0}+x_{2,1,0}$ in 
Eq.(\ref{totalsuperpotential}). We get the following 
superpotential
\begin{eqnarray}
W^{\prime}_{break}&=&(mx_{0}+fx^{2}_{1})x_{2}+(m\hat{S}_{0}+2fx_{1}\hat{S}_{1})x_{2}+
(mx_{0}+fx^{2}_{1}) \hat{S}_{2}+(m \hat{S}_{0}+2fx_{1}\hat{S}_{1}) \hat{S}_{2}+f\hat{S}_{1}\hat{S}_{1}\hat{S}_{2}
 \nonumber \\
&+&fx_{2}\hat{S}_{1}\hat{S}_{1}+h^{M}x_{1}\hat{N}^{c}\hat{N}^{c}+h^{H}x_{2}\hat{H}_{1}\hat{H}_{2}+ 
h^{M}\hat{S}_{1}\hat{N}^{c}\hat{N}^{c}+h^{H}\hat{S}_{2}\hat{H}_{1}\hat{H}_{2}.  
\end{eqnarray}
We want to emphasaise, the 
$\mu \equiv h^{H}x_{2}$, therefore, $x_{2}$ must be of order $M_{SUSY}$ and
\begin{equation}
mx_{0}+fx^{2}_{1}={\cal O}(M_{SUSY}).
\end{equation}
We first break Peccei-Quinn symmetry and later, but at same energy scale, we broke supersymmetry.

If we use the Eq.(\ref{getaxion}) in the first line in equation above, we get the following 
result
\begin{equation}
\frac{m}{x_{1}}\left( x_{1}\hat{S}_{0}-2x_{0}\hat{S}_{1}\right)\hat{S}_{2}+
f\hat{S}_{1}\hat{S}_{1}\hat{S}_{2},
\end{equation}
therefore the linear combination
\begin{equation}
\hat{A}\propto 2x_{0}\hat{S}_{0}+x_{1}\hat{S}_{1},
\end{equation}
is a massless superfield. As $x_{1}\gg x_{0}$ the axion 
superfield is almost given by $\hat{S}_{1}$.

\section{Conclusion}
\label{sec:conclusion} 

We start with a supersymmetric theory of just one large fundamental mass $m$. We assume it to 
be invariant under an anomalous global U(1) symmetry to include three singlet (right-handed) 
neutrino superfields $(\hat{N}^{c})_{i}$ and three other singlet superfields 
$\hat{S}_{0,1,2}$. The supersymmetry is then softly broken at $M_{SUSY}$ of order 1 TeV.  
As a result of the assumed particle content of the theory, an axion scale $f_{a}$ 
of order $m$ is generated, from which neutrinos obtain masses 
via the usual seesaw mechanism with $m_{N} \sim f_{a}$. 

\begin{center}
{\bf Acknowledgments} 
\end{center}
The author would like to thanks to Instituto de F\'\i sica 
Te\'orica (IFT-Unesp) for their nice hospitality during the 
period I developed this work was accomplished.


\end{document}